# Multi-Domain Collaborative Filtering

**Yu Zhang, Bin Cao, Dit-Yan Yeung**
Department of Computer Science and Engineering, Hong Kong University of Science and Technology
Clear Water Bay, Kowloon, Hong Kong, China
{zhangyu,caobin,dyyeung}@cse.ust.hk

## Abstract

Collaborative filtering is an effective recommendation approach in which the preference of a user on an item is predicted based on the preferences of other users with similar interests. A big challenge in using collaborative filtering methods is the data sparsity problem which often arises because each user typically only rates very few items and hence the rating matrix is extremely sparse. In this paper, we address this problem by considering multiple collaborative filtering tasks in different domains simultaneously and exploiting the relationships between domains. We refer to it as a multi-domain collaborative filtering (MCF) problem. To solve the MCF problem, we propose a probabilistic framework which uses probabilistic matrix factorization to model the rating problem in each domain and allows the knowledge to be adaptively transferred across different domains by automatically learning the correlation between domains. We also introduce the link function for different domains to correct their biases. Experiments conducted on several real-world applications demonstrate the effectiveness of our methods when compared with some representative methods.

## 1 Introduction

The amount of information available on the Internet is increasing at an astonishing rate, making information search a more and more challenging task. As such, recommendation plays an important role to bring items of potential interest to our attention. Some popular examples include systems for product recommendation in Amazon.com, movie recommendation in Netflix and MovieLens, and reference recommendation in CiteULike. Collaborative filtering (CF) is an effective recommendation approach based on the intuitive idea that the preference of a user can be predicted by exploiting the information about other users which share similar interests. In particular, CF techniques exploit past activities of the users, such as their transaction history or product satisfaction expressed in ratings, to predict the future activities of the users. In recent years, CF-based recommendation systems have become increasingly popular because it is generally much easier to collect the past activities of users than their profiles, partially due to privacy considerations.

According to a survey on CF [26], different CF techniques can be classified into three categories: memory-based methods, model-based methods, and hybrid methods. Similar to the idea of nearest neighbor classification, memory-based methods make rating prediction based on the rating behavior of other items and users with similar interests. Some representative methods are [10, 23, 16]. One limitation of memory-based methods is that they require the rating data to be dense so that the similarity values can be estimated accurately. Unfortunately, this requirement is not realistic in many applications. To achieve better prediction performance and overcome the shortcomings of memory-based CF methods, model-based CF methods have been proposed and actively studied. Model-based CF techniques use the rating data to learn a model and then use the learned model to make predictions. Many learning models have been used for CF, such as Bayesian belief networks [4], graphical models [11, 28], and dependency networks [9]. Among all model-based CF methods, matrix factorization methods are perhaps the most popular one in recent years [2, 25, 20, 5, 21, 22, 14, 32, 29]. These methods assume that the user and item features lie in some low-dimensional latent space and then make predictions based on the latent features. With the hope to improve performance further, hybrid CF techniques have been proposed to combine memory-based methods with model-based methods, or utilize additional information such as content information. Some examples are [1, 18, 19, 17, 30, 31, 13].

Even though CF methods have achieved great successes in recommendation applications, there still exist some problems which limit their performance. A big challenge is the data sparsity problem [26] which means that the rating matrix is extremely sparse. Our focus in this paper is on this data sparsity problem. In particular, we consider a multi-domain CF (MCF) problem which jointly models a collection of rating prediction tasks arising from multiple domains. The MCF problem is particularly suitable for large-scale e-commerce and social networking services which often provide a diverse range of products or services. For example, different product or service categories such as books and electronics naturally constitute different domains. By exploiting the correlation between rating prediction problems in different domains, we can transfer the shared knowledge among similar domains to alleviate the data sparsity problem and therefore improve the rating prediction performance in all domains. Specifically, we propose a probabilistic framework which uses probabilistic matrix factorization (PMF) [21] to model the rating prediction problem in each domain and allows the knowledge to be adaptively transferred across different domains by automatically learning the correlation between domains. We also introduce the link function for different domains to correct their biases. Experiments conducted on several real-world applications demonstrate the effectiveness of our method.

The rest of this paper is organized as follows. We present our method in Section 2 and some improvement in Section 3. Some related works are discussed in Section 4. Section 5 reports the experimental results based on some recommendation datasets. Concluding remarks are given in the final section.

## 2 Multi-Domain Collaborative Filtering

Let $\mathbf{X}^i \in \mathbb{R}^{m_i \times n_i}$ denote the rating matrix for the $i$th domain, where $i = 1, \ldots, K$. So for each domain we have $m_i$ users and $n_i$ items. In total we have $m$ users in all domains. Let $\mathbf{U}^i \in \mathbf{R}^{d \times m}$ and $\mathbf{V}^i \in \mathbb{R}^{d \times n_i}$ denote the latent user and item feature matrices with each column $\mathbf{U}^i_j$ and each column $\mathbf{V}^i_k$ representing the user-specific and item-specific latent feature vectors, respectively.

We define the conditional distribution over the observed ratings on the $i$th domain as

$$p(\mathbf{X}^i|\mathbf{U}^i, \mathbf{V}^i, \sigma_i) = \prod_{j=1}^{m}\prod_{k=1}^{n_i}\left[\mathcal{N}(X^i_{jk}|(\mathbf{U}^i_j)^T\mathbf{V}^i_k, \sigma_i^2)\right]^{I^i_{jk}}, \quad (1)$$

where $\mathcal{N}(\mathbf{m}, \boldsymbol{\Sigma})$ denotes the multivariate (or univariate) normal distribution with mean $\mathbf{m}$ and covariance matrix (or variance) $\boldsymbol{\Sigma}$, $X^i_{jk}$ denotes the rating of the $j$th user on the $k$th element in $\mathbf{X}^i$, and $I^i_{jk}$ is the indicator function which is equal to 1 if the $j$th user rated the $k$th item in the $i$th domain and is 0 otherwise.

We place zero-mean spherical Gaussian priors [27] on the user features and item features as

$$p(\mathbf{U}^i|\lambda_i) = \prod_{j=1}^{m}\mathcal{N}(\mathbf{U}^i_j|\mathbf{0}_d, \lambda_i^2\mathbf{I}_d) \quad (2)$$

$$p(\mathbf{V}^i|\eta_i) = \prod_{k=1}^{n_i}\mathcal{N}(\mathbf{V}^i_k|\mathbf{0}_d, \eta_i^2\mathbf{I}_d), \quad (3)$$

where $\mathbf{0}_d$ denotes the $d \times 1$ zero vector and $\mathbf{I}_d$ denotes the $d \times d$ identity matrix.

To learn the relationships between different domains, we place a matrix-variate normal distribution [8] on $\mathbf{U} = [\text{vec}(\mathbf{U}^1), \ldots, \text{vec}(\mathbf{U}^K)]$ where $\text{vec}(\cdot)$ denotes the operator which converts a matrix into a vector in a columnwise manner:

$$p(\mathbf{U}|\boldsymbol{\Omega}) = \mathcal{MN}_{md \times K}(\mathbf{U}|\mathbf{0}_{md \times K}, \mathbf{I}_{md} \otimes \boldsymbol{\Omega}), \quad (4)$$

where $\mathbf{0}_{a \times b}$ denotes an $a \times b$ zero matrix, and $\mathcal{MN}_{a \times b}(\mathbf{M}, \mathbf{A} \otimes \mathbf{B})$ denotes a matrix-variate normal distribution with mean $\mathbf{M} \in \mathbb{R}^{a \times b}$, row covariance matrix $\mathbf{A} \in \mathbb{R}^{a \times a}$ and column covariance matrix $\mathbf{B} \in \mathbb{R}^{b \times b}$. The probability density function of the matrix-variate normal distribution is defined as

$$p(\mathbf{X}|\mathbf{M}, \mathbf{A}, \mathbf{B}) = \frac{\exp\left(-\frac{1}{2}\text{tr}\left(\mathbf{A}^{-1}(\mathbf{X} - \mathbf{M})\mathbf{B}^{-1}(\mathbf{X} - \mathbf{M})^T\right)\right)}{(2\pi)^{ab/2}|\mathbf{A}|^{b/2}|\mathbf{B}|^{a/2}},$$

where $\text{tr}(\cdot)$ and $|\cdot|$ denote the trace and determinant, respectively, of a square matrix. More specifically, here the row covariance matrix $\mathbf{I}_{md}$ models the relationships between user latent features and the column covariance matrix $\boldsymbol{\Omega}$ models the relationships between different $\mathbf{U}^i$'s. In other words, $\boldsymbol{\Omega}$ models the relationships between domains.

### 2.1 Parameter Learning

The log-posterior over $\{\mathbf{U}^i\}$ and $\{\mathbf{V}^i\}$ is given by

$$\ln p(\{\mathbf{U}^i\}, \{\mathbf{V}^i\}|\{\mathbf{X}^i\}, \boldsymbol{\sigma}, \boldsymbol{\lambda}, \boldsymbol{\eta}, \boldsymbol{\Omega})$$
$$= -\sum_{i=1}^{K}\frac{1}{2\sigma_i^2}\sum_{j=1}^{m}\sum_{k=1}^{n_i}I^i_{jk}\left(X^i_{jk} - (\mathbf{U}^i_j)^T\mathbf{V}^i_k\right)^2$$
$$- \sum_{i=1}^{K}\frac{1}{2\lambda_i^2}\sum_{j=1}^{m}(\mathbf{U}^i_j)^T\mathbf{U}^i_j - \sum_{i=1}^{K}\frac{1}{2\eta_i^2}\sum_{k=1}^{n_i}(\mathbf{V}^i_k)^T\mathbf{V}^i_k$$
$$- \frac{1}{2}\sum_{i=1}^{K}(\ln\sigma_i^2\sum_{j=1}^{m}\sum_{k=1}^{n_i}I^i_{jk}) - \frac{md}{2}\sum_{i=1}^{K}\ln\lambda_i^2$$
$$- \sum_{i=1}^{K}\frac{dn_i}{2}\ln\eta_i^2 - \frac{1}{2}\text{tr}(\mathbf{U}\boldsymbol{\Omega}^{-1}\mathbf{U}^T) - \frac{md}{2}\ln|\boldsymbol{\Omega}| + \text{Const},$$
$$(5)$$

where $\boldsymbol{\sigma} = (\sigma_1, \ldots, \sigma_K)^T$, $\boldsymbol{\lambda} = (\lambda_1, \ldots, \lambda_K)^T$, and $\boldsymbol{\eta} = (\eta_1, \ldots, \eta_K)^T$. We maximize $\ln p(\{\mathbf{U}^i\}, \{\mathbf{V}^i\}|\{\mathbf{X}^i\}, \boldsymbol{\sigma}, \boldsymbol{\lambda}, \boldsymbol{\eta}, \boldsymbol{\Omega})$ to obtain the *maximum a posteriori* (MAP) solution of $\{\mathbf{U}^i\}$ and $\{\mathbf{V}^i\}$ and the *maximum likelihood estimation* (MLE) solution of $\boldsymbol{\sigma}$, $\boldsymbol{\lambda}$, $\boldsymbol{\eta}$ and $\boldsymbol{\Omega}$. We use an alternating method to minimize $J(\{\mathbf{U}^i\}, \{\mathbf{V}^i\}, \boldsymbol{\sigma}, \boldsymbol{\lambda}, \boldsymbol{\eta}, \boldsymbol{\Omega}) = -\ln p(\{\mathbf{U}^i\}, \{\mathbf{V}^i\}|\{\mathbf{X}^i\}, \boldsymbol{\sigma}, \boldsymbol{\lambda}, \boldsymbol{\eta}, \boldsymbol{\Omega})$. In what follows, we will present each subproblem separately.

**Optimizing w.r.t. $\mathbf{U}_j^i$ when the other variables are fixed**

The derivative of $J$ with respect to $\mathbf{U}_j^i$ can be calculated as

$$\frac{\partial J}{\partial \mathbf{U}_j^i} = \frac{1}{\lambda_i^2}\mathbf{U}_j^i + \frac{1}{\sigma_i^2}\sum_{k=1}^{n_i} I_{jk}^i \Big(\mathbf{V}_k^i(\mathbf{V}_k^i)^T \mathbf{U}_j^i - X_{jk}^i \mathbf{V}_k^i\Big) + \sum_{l=1}^{K} \mathbf{U}_j^l \psi_{li},$$

where $\boldsymbol{\Psi} = \boldsymbol{\Omega}^{-1}$ and $\psi_{ij}$ is the $(i,j)$th element of $\boldsymbol{\Psi}$. We set the derivative to zero and obtain the analytical solution as

$$\mathbf{U}_j^i = \Big(\sigma_i^2(\frac{1}{\lambda_i^2} + \psi_{ii})\mathbf{I}_d + \sum_{k=1}^{n_i} I_{jk}^i \mathbf{V}_k^i(\mathbf{V}_k^i)^T\Big)^{-1} \cdot$$
$$\Big(\sum_{k=1}^{n_i} I_{jk}^i X_{jk}^i \mathbf{V}_k^i - \sigma_i^2 \sum_{l\neq i} \psi_{li} \mathbf{U}_j^l\Big). \quad (6)$$

Consider a special case in which different domains are uncorrelated, which means that $\boldsymbol{\Omega}$ and $\boldsymbol{\Psi}$ are diagonal matrices, i.e., $\Psi_{ij} = 0$ for $i \neq j$. Then the update solution for $\mathbf{U}_j^i$ is

$$\mathbf{U}_j^i = \Big(\sigma_i^2(\frac{1}{\lambda_i^2} + \psi_{ii})\mathbf{I}_d + \sum_{k=1}^{n_i} I_{jk}^i \mathbf{V}_k^i(\mathbf{V}_k^i)^T\Big)^{-1} \sum_{k=1}^{n_i} I_{jk}^i X_{jk}^i \mathbf{V}_k^i,$$

which degenerates to the update solution for single-domain matrix factorization.

**Optimizing w.r.t. $\mathbf{V}_k^i$ when the other variables are fixed**

The derivative of $J$ with respect to $\mathbf{V}_k^i$ can be calculated as

$$\frac{\partial J}{\partial \mathbf{V}_k^i} = \frac{1}{\eta_i^2}\mathbf{V}_k^i + \frac{1}{\sigma_i^2}\sum_{j=1}^{m} I_{jk}^i \Big(\mathbf{U}_j^i(\mathbf{U}_j^i)^T \mathbf{V}_k^i - X_{jk}^i \mathbf{U}_j^i\Big).$$

We set the derivative to zero and obtain the analytical solution as

$$\mathbf{V}_k^i = \Big(\frac{\sigma_i^2}{\eta_i^2}\mathbf{I}_d + \sum_{j=1}^{m} I_{jk}^i \mathbf{U}_j^i(\mathbf{U}_j^i)^T\Big)^{-1} \sum_{j=1}^{m} I_{jk}^i X_{jk}^i \mathbf{U}_j^i. \quad (7)$$

**Optimizing w.r.t. $\boldsymbol{\Omega}$ when the other variables are fixed**

Since $\boldsymbol{\Omega}$ is defined as a covariance matrix, $\boldsymbol{\Omega}$ and $\boldsymbol{\Omega}^{-1}$ are symmetric matrices. Then the derivative of $J$ with respect to $\boldsymbol{\Omega}^{-1}$ can be calculated as

$$\frac{\partial J}{\partial \boldsymbol{\Omega}^{-1}} = \mathbf{U}^T\mathbf{U} - md\,\boldsymbol{\Omega} - \frac{1}{2}(\mathbf{U}^T\mathbf{U} - md\,\boldsymbol{\Omega}) \odot \mathbf{I}_K,$$

where $\odot$ denotes the Hadamard product which is the matrix elementwise product. We set the derivative to zero and get

$$\boldsymbol{\Sigma} = \frac{1}{2}(\boldsymbol{\Sigma} \odot \mathbf{I}_K),$$

where $\boldsymbol{\Sigma} = \mathbf{U}^T\mathbf{U} - md\,\boldsymbol{\Omega}$. Then we have

$$\Sigma_{ii} = \Sigma_{ii}/2 \Rightarrow \Sigma_{ii} = 0$$
$$\Sigma_{ij} = 0, i \neq j,$$

where $\Sigma_{ij}$ is the $(i,j)$th element of $\boldsymbol{\Sigma}$. So $\boldsymbol{\Sigma}$ is a zero matrix and we obtain the analytical solution for $\boldsymbol{\Omega}$ as

$$\boldsymbol{\Omega} = \frac{1}{md}\mathbf{U}^T\mathbf{U}. \quad (8)$$

Considering Eq. (8), the $(i,j)$th element $\Omega_{ij}$ of $\boldsymbol{\Omega}$, which corresponds to the covariance between the $i$th and $j$th domains, can be computed as

$$\Omega_{ij} = \frac{1}{md}\Big(\mathrm{vec}(\mathbf{U}^i)\Big)^T \mathrm{vec}(\mathbf{U}^j),$$

which is the scaled dot product of $\mathrm{vec}(\mathbf{U}^i)$ and $\mathrm{vec}(\mathbf{U}^j)$. Since $\mathrm{vec}(\mathbf{U}^i)$ is modeled as latent user features in the $i$th domain, using the dot product to represent covariance matches our intuition.

**Optimizing w.r.t. $\sigma_i$ when the other variables are fixed**

The derivative of $J$ with respect to $\sigma_i^2$ can be calculated as

$$\frac{\partial J}{\partial \sigma_i^2} = -\frac{1}{2\sigma_i^4}\sum_{j=1}^{m}\sum_{k=1}^{n_i} I_{jk}^i \Big(X_{jk}^i - (\mathbf{U}_j^i)^T\mathbf{V}_k^i\Big)^2 + \frac{1}{2\sigma_i^2}\sum_{j=1}^{m}\sum_{k=1}^{n_i} I_{jk}^i.$$

We set the derivative to zero and obtain the analytical solution as

$$\sigma_i^2 = \frac{\sum_{j=1}^{m}\sum_{k=1}^{n_i} I_{jk}^i \Big(X_{jk}^i - (\mathbf{U}_j^i)^T\mathbf{V}_k^i\Big)^2}{\sum_{j=1}^{m}\sum_{k=1}^{n_i} I_{jk}^i}. \quad (9)$$

**Optimizing w.r.t. $\lambda_i$ when the other variables are fixed**

The derivative of $J$ with respect to $\lambda_i^2$ can be calculated as

$$\frac{\partial J}{\partial \lambda_i^2} = -\frac{1}{2\lambda_i^4}\sum_{j=1}^{m}(\mathbf{U}_j^i)^T\mathbf{U}_j^i + \frac{md}{2\lambda_i^2}.$$

We set the derivative to zero and obtain the analytical solution as

$$\lambda_i^2 = \frac{1}{md}\sum_{j=1}^{m}(\mathbf{U}_j^i)^T\mathbf{U}_j^i. \quad (10)$$

**Optimizing w.r.t. $\eta_i$ when the other variables are fixed**

The derivative of $J$ with respect to $\eta_i^2$ can be calculated as

$$\frac{\partial J}{\partial \eta_i^2} = -\frac{1}{2\eta_i^4}\sum_{k=1}^{n_i}(\mathbf{V}_k^i)^T\mathbf{V}_k^i + \frac{dn_i}{2\eta_i^2}.$$

We set the derivative to zero and obtain the analytical solution as

$$\eta_i^2 = \frac{1}{dn_i}\sum_{k=1}^{n_i}(\mathbf{V}_k^i)^T\mathbf{V}_k^i. \quad (11)$$

## 2.2 Discussions

To gain more insights into our method, we plug Eqs. (8), (10) and (11) into $J(\{\mathbf{U}^i\}, \{\mathbf{V}^i\}, \boldsymbol{\sigma}, \boldsymbol{\lambda}, \boldsymbol{\eta}, \boldsymbol{\Omega})$. By ignoring some constant terms, $J(\{\mathbf{U}^i\}, \{\mathbf{V}^i\}, \boldsymbol{\sigma}, \boldsymbol{\lambda}, \boldsymbol{\eta}, \boldsymbol{\Omega})$ can be reformulated as

$$
\begin{aligned}
&J(\{\mathbf{U}^i\}, \{\mathbf{V}^i\}, \boldsymbol{\sigma}, \boldsymbol{\lambda}, \boldsymbol{\eta}, \boldsymbol{\Omega}) \\
&= \sum_{i=1}^K \frac{1}{2\sigma_i^2} \sum_{j=1}^m \sum_{k=1}^{n_i} I_{jk}^i \Big(X_{jk}^i - (\mathbf{U}_j^i)^T \mathbf{V}_k^i\Big)^2 \\
&+ \frac{1}{2} \sum_{i=1}^K (\ln \sigma_i^2 \sum_{j=1}^m \sum_{k=1}^{n_i} I_{jk}^i) + \frac{md}{2} \sum_{i=1}^K \ln \Big( \sum_{j=1}^m (\mathbf{U}_j^i)^T \mathbf{U}_j^i \Big) \\
&+ \sum_{i=1}^K \frac{dn_i}{2} \ln \Big( \sum_{k=1}^{n_i} (\mathbf{V}_k^i)^T \mathbf{V}_k^i \Big) + \frac{1}{2(md)^{K-1}} \ln |\mathbf{U}^T \mathbf{U}|.
\end{aligned} \quad (12)
$$

The first term in Eq. (12) measures the empirical loss on the observed ratings, the second term penalizes the complexity of $\boldsymbol{\sigma}$, the third and fifth terms penalize the complexity of $\{\mathbf{U}^i\}$, and the fourth term penalizes the complexity of $\{\mathbf{V}^i\}$.

Since

$$
\ln \Big( \sum_{j=1}^m (\mathbf{U}_j^i)^T \mathbf{U}_j^i \Big) = \ln \operatorname{tr}\Big(\mathbf{U}^i (\mathbf{U}^i)^T\Big)
$$

$$
\ln \Big( \sum_{k=1}^{n_i} (\mathbf{V}_k^i)^T \mathbf{V}_k^i \Big) = \ln \operatorname{tr}\Big(\mathbf{V}^i (\mathbf{V}^i)^T\Big),
$$

which are related to the trace norms of $\mathbf{U}^i$ and $\mathbf{V}^i$ [25], the third and fourth terms in Eq. (12) penalize the ranks of $\mathbf{U}^i$ and $\mathbf{V}^i$, respectively [6]. Moreover, according to [7], using the last term in Eq. (12) is to minimize the product of all singular values of $\mathbf{U}^i$ which is related to the rank of $\mathbf{U}^i$.

## 3 Incorporation of Link Function

In the above model, the likelihood for the ratings is Gaussian as defined in Eq. (1). However, since the ratings are discrete integral values, Gaussian likelihood is not very suitable and hence using it may affect the performance of our model. Here we consider a modification of our model which first transforms the original ratings by a so-called link function and then applies the above model on the transformed ratings. In what follows, we will present this modification in detail.

The link function for the $i$th domain is denoted by $g_i(\cdot; \boldsymbol{\theta}_i)$ which is parameterized by $\boldsymbol{\theta}_i$. We require $g_i$ to be monotonically increasing and mapping onto the whole real line; otherwise the probability measure will not be preserved after the transformation. The transformed ratings are denoted by latent variables $Z_{jk}^i = g_i(X_{jk}^i)$. Similar to Eq. (1), the likelihood is defined on $Z_{jk}^i$ as

$$
p(\mathbf{Z}^i | \mathbf{U}^i, \mathbf{V}^i, \sigma_i) = \prod_{j=1}^m \prod_{k=1}^{n_i} \Big[ \mathcal{N}(Z_{jk}^i | (\mathbf{U}_j^i)^T \mathbf{V}_k^i, \sigma_i^2) \Big]^{I_{jk}^i}.
$$

Then, using the Jacobian transformation, we obtain the likelihood on $\mathbf{X}^i$ as

$$
\begin{aligned}
&p(\mathbf{X}^i | \mathbf{U}^i, \mathbf{V}^i, \sigma_i) \\
&= \prod_{j=1}^m \prod_{k=1}^{n_i} \Big[ \mathcal{N}\Big(g_i(X_{jk}^i) | (\mathbf{U}_j^i)^T \mathbf{V}_k^i, \sigma_i^2\Big) g_i'(X_{jk}^i) \Big]^{I_{jk}^i},
\end{aligned} \quad (13)
$$

where $g_i'(\cdot)$ denotes the derivative function of $g_i(\cdot)$. For simplicity of discussion, we assume that different domains share the same link function, i.e., $g_i(\cdot) = g_j(\cdot), \forall i \neq j$. We denote the common link function as $g(\cdot)$ which is parameterized by $\boldsymbol{\theta}$.

For parameter learning, we still maximize the log-posterior to get the MAP solution of $\{\mathbf{U}^i\}$ and $\{\mathbf{V}^i\}$ and the MLE solution of the model parameters including $\boldsymbol{\sigma}, \boldsymbol{\lambda}, \boldsymbol{\eta}, \boldsymbol{\Omega}$ and $\boldsymbol{\theta}$. In this way, both the original model parameters in the above section and the parameters of the link function are learned simultaneously under the same probabilistic framework. We still use an alternating method to optimize the objective function.

In detail, the negative log-posterior of the whole data is computed as

$$
\begin{aligned}
&J_1(\{\mathbf{U}^i\}, \{\mathbf{V}^i\}, \boldsymbol{\sigma}, \boldsymbol{\lambda}, \boldsymbol{\eta}, \boldsymbol{\Omega}, \boldsymbol{\theta}) \\
&= \sum_{i=1}^K \frac{1}{2\sigma_i^2} \sum_{j=1}^m \sum_{k=1}^{n_i} I_{jk}^i \Big(g(X_{jk}^i) - (\mathbf{U}_j^i)^T \mathbf{V}_k^i\Big)^2 \\
&+ \sum_{i=1}^K \frac{1}{2\lambda_i^2} \sum_{j=1}^m (\mathbf{U}_j^i)^T \mathbf{U}_j^i + \sum_{i=1}^K \frac{1}{2\eta_i^2} \sum_{k=1}^{n_i} (\mathbf{V}_k^i)^T \mathbf{V}_k^i \\
&+ \frac{1}{2} \sum_{i=1}^K (\ln \sigma_i^2 \sum_{j=1}^m \sum_{k=1}^{n_i} I_{jk}^i) + \frac{md}{2} \sum_{i=1}^K \ln \lambda_i^2 \\
&+ \sum_{i=1}^K \frac{dn_i}{2} \ln \eta_i^2 + \frac{1}{2} \operatorname{tr}(\mathbf{U} \boldsymbol{\Omega}^{-1} \mathbf{U}^T) + \frac{md}{2} \ln |\boldsymbol{\Omega}| \\
&- \sum_{i=1}^K \sum_{j=1}^m \sum_{k=1}^{n_i} I_{jk}^i \ln g'(X_{jk}^i) + \text{Const.}
\end{aligned} \quad (14)
$$

The update equations for $\{\mathbf{U}^i\}, \{\mathbf{V}^i\}, \boldsymbol{\sigma}, \boldsymbol{\lambda}$ and $\boldsymbol{\eta}$ are similar to Eqs. (6)–(11) by replacing $X_{jk}^i$ with $g(X_{jk}^i)$. For the learning of $\boldsymbol{\theta}$, since there is no analytical update solution, we use a gradient-based method such as the scaled conjugate gradient method[1] to update $\boldsymbol{\theta}$. More specifically, the gradient of $J_1$ with respect to $\theta_l$, the $l$th element of $\boldsymbol{\theta}$, is

---

[1] http://www.kyb.tuebingen.mpg.de/bs/people/carl/code/minimize/minimize.m

computed as

$$\frac{\partial J_1}{\partial \theta_l} = \sum_{i=1}^{K} \frac{1}{\sigma_i^2} \sum_{j=1}^{m} \sum_{k=1}^{n_i} I_{jk}^i \Big(g(X_{jk}^i) - (\mathbf{U}_j^i)^T \mathbf{V}_k^i\Big) \frac{\partial g(X_{jk}^i)}{\partial \theta_l}$$
$$- \sum_{i=1}^{K} \sum_{j=1}^{m} \sum_{k=1}^{n_i} I_{jk}^i \frac{\partial \ln g'(X_{jk}^i)}{\partial \theta_l}.$$

When we want to predict the $(q, r)$th element in $\mathbf{X}^i$, we first predict the latent variable $Z_{qr}^i$ as

$$\tilde{Z}_{qr}^i = (\mathbf{U}_q^i)^T \mathbf{V}_r^i,$$

and then the prediction for $X_{qr}^i$ is computed as

$$\tilde{X}_{qr}^i = g^{-1}(\tilde{Z}_{qr}^i),$$

where $g^{-1}(\cdot)$ denotes the inverse function of $g(\cdot)$. When $g(\cdot)$ has a simple form, we can easily find the form of $g^{-1}(\cdot)$; when $g^{-1}(\cdot)$ is not easy to obtain, we can use numerical methods such as the bisection method to find the unique root of the equation $g(x) = \tilde{Z}_{qr}^i$ due to the monotonic property of $g(\cdot)$.

In our experiments, we use $g(x) = a \ln(bx + c) + d$ $(a, b, c > 0, d \in \mathbb{R})$ as the link function. Here $g(x)$ is monotonically increasing and its domain is the set of all real numbers.

## 4 Related Work

There is not much previous work on the MCF problem. The most related one is [24] which proposes a collective matrix factorization (CMF) method for CF. For the case of MCF, the collective matrix factorization method requires a common latent user feature matrix $\mathbf{U}$ which is shared by all domains. However, in real applications in which different domains have heterogenous properties, this requirement is not very reasonable. Our model can be viewed as a generalization of the collective matrix factorization method where each domain has its own latent user feature matrix and the correlation matrix between different domains is learned to improve the performance of all rating problems in all domains. In this sense, collective matrix factorization can be viewed as a special case of our model by restricting all $\mathbf{U}^i$ to be identical. Moreover, a transfer collaborative filtering model was proposed in [15] which aims at improving the performance of a rating problem with very sparse data with the help of another rating problem which has denser rating data. However, the objective of [15] is different from ours. For example, the model in [15] is to improve one rating problem with the help of another rating problem, but in our case, we want to improve the performance of all rating problems in all domains simultaneously. Moreover, the model in [15] seems to work only for problems with two domains while our model can work for two or more domains in the same way.

## 5 Experiments

In this section, we report some experiments on two real-world datasets along with our analysis on the results.

### 5.1 Experimental Settings

We test the proposed methods on two public-domain recommendation datasets in which the items come from different domains or sub-domains.

#### 5.1.1 Datasets

We use two commonly used datasets in our experiments including one from movie ratings and one from book ratings. In both datasets, the items can be divided into multiple heterogeneous domains.

- MovieLens[2] is a widely used movie recommendation dataset. It contains 100,000 ratings in the scale 1–5. The ratings are given by 943 users on 1,682 movies. Besides the rating information, genre information about movies is also available.

- Book-Crossing[3] is a public book ratings dataset. A subset of the data is used in our experiment, consisting of ratings on books with category information available on Amazon.com. The subset contains 56,148 ratings in the scale 1–10 and these ratings are given by 28,503 users on 9,009 books.

For the MovieLens dataset, we use the five most popular genres to define the domains, whereas for the Book-Crossing dataset, we use the five general book categories. We randomly select 80% of the rating data from each domain to form the training set and the rest for the test set. Each configuration is iterated 10 times in the experiments.

#### 5.1.2 Evaluation Metric

In this paper, we use root mean squared error (RMSE) as the measure for performance evaluation:

$$\text{RMSE} = \sqrt{\frac{\sum_{i,j}(r_{ij} - \hat{r}_{ij})^2}{N}}, \quad (15)$$

where $r_{ij}$ denotes the ground-truth rating of user $i$ for item $j$, $\hat{r}_{ij}$ denotes the predicted rating, and the denominator $N$ is the number of ratings tested. The smaller the RMSE score, the better the performance.

#### 5.1.3 Baselines

We compare our proposed models with the following two methods:

---
[2] http://www.grouplens.org/
[3] http://www.informatik.uni-freiburg.de/~cziegler/BX/

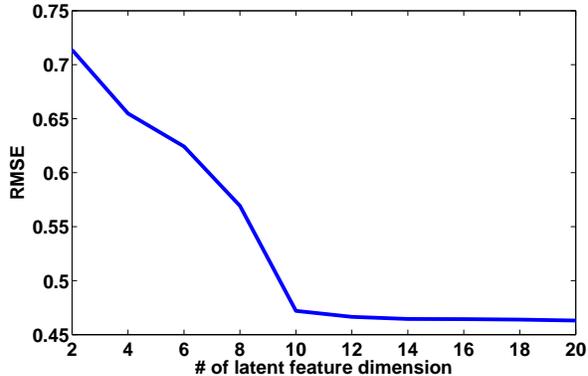

Figure 1: Effect of the latent feature dimensionality on the performance of rating prediction for a subset of the MovieLens data.

- Independent collaborative filtering using probabilistic matrix factorization (PMF), which treats different rating prediction problems in different domains independently.

- Collective matrix factorization (CMF) model [24], which handles problems involving multiple matrix factorization tasks.

In the following, we refer to our proposed method in Section 2 as MCF and the one in Section 3 as MCF-LF.

## 5.2 Experimental Results

### 5.2.1 Parameter Setting

An appealing advantage of our probabilistic model is that it has very few parameters to set. In fact, the only parameter that needs to be set is the latent dimensionality $d$. Figure 1 shows the effect of the latent dimensionality on the performance of MCF for a subset of the MovieLens dataset. We can see that the performance in terms of RMSE does not change much after $d$ reaches 10. Therefore, we set $d$ to 10 in the following experiments. Other parameters in PMF, CMF, MCF, MCF-LF are randomly generated.

### 5.2.2 Results

Table 1 shows the experimental results on the MovieLens dataset. We can see that our proposed models have the best performance. The models that take multiple domains into consideration (CMF, MCF, MCF-LF) perform better than PMF which treats different domains independently. MCF, which can learn the similarities between different rating prediction problems, performs better than CMF, demonstraing the effectiveness of exploiting the relationships between different domains. Comparing MCF with its variant MCF-LF which has the link function, we can conclude that the link function brings about performance improvement consistently over all domains.

Table 2 shows the experimental results on the Book-Crossing dataset. MCF and MCF-LF are also the best among all methods compared. Different from the situation in the MovieLens dataset, the performance of CMF is worse than that of PMF, even though CMF considers multiple domains jointly. The reason can be inferred from the correlation matrix in Table 4. Since some domains are uncorrelated (1st and 4th domains, and 2nd and 4th domains), the assumption in CMF that different domains share the same latent user features seems not very reasonable, making the performance of CMF worse than that of PMF. However, since our methods can take the correlations between different domains into consideration, they can achieve better performance.

### 5.2.3 Analysis on Correlation Matrix

Table 3 shows the correlation matrix between five domains learned from the MovieLens dataset, which seems consistent with intuition. For example, the genres 'Comedy' and 'Thriller' have the smallest correlation while 'Romance' and 'Drama' have the largest one. For the genre 'Action', we can see that the other genres are ranked into order as: 'Thriller', 'Romance', 'Drama', and 'Comedy', which matches our intuition.

Table 4 shows the correlation matrix between five domains learned from the Book-Crossing dataset. Some relations between different domains revealed also seem intuitive. For example, the categories 'Mystery & Thrillers' and 'Business & Investing' have nearly zero correlation and the same is true for 'Science Fiction & Fantasy' and 'Business & Investing'. Also, categories 'Science' and 'Religion & Spirituality' have the largest correlation.

Table 3: Mean of correlation matrix learned by MCF-LF on the MovieLens data in different domains. 1st domain: 'Comedy'; 2nd domain: 'Romance'; 3rd domain: 'Drama'; 4th domain: 'Action'; 5th domain: 'Thriller'.

|     | 1st    | 2nd    | 3rd    | 4th    | 5th    |
|-----|--------|--------|--------|--------|--------|
| 1st | 1.0000 | 0.8837 | 0.8584 | 0.8319 | 0.8302 |
| 2nd | 0.8837 | 1.0000 | 0.9288 | 0.8855 | 0.8805 |
| 3rd | 0.8584 | 0.9288 | 1.0000 | 0.8647 | 0.8783 |
| 4th | 0.8319 | 0.8855 | 0.8647 | 1.0000 | 0.9122 |
| 5th | 0.8302 | 0.8805 | 0.8783 | 0.9122 | 1.0000 |

## 6 Conclusion

In this paper, we have addressed the multi-domain collaborative filtering problem in which multiple rating prediction problems are jointly learned. We propose a probabilis-tic model which considers the correlation between different domains when leveraging all rating data together. Experiments conducted on several recommendation datasets demonstrate the effectiveness of our methods.

Table 1: Comparison of different methods on the MovieLens data. Each column records the RMSE scores on one domain and the last column records the RMSE score on the total testing data. Each row records the mean RMSE of the corresponding method over 10 trials. 1st domain: 'Comedy'; 2nd domain: 'Romance'; 3rd domain: 'Drama'; 4th domain: 'Action'; 5th domain: 'Thriller'.

| Method | 1st domain | 2nd domain | 3rd domain | 4th domain | 5th domain | Total |
| --- | --- | --- | --- | --- | --- | --- |
| PMF | 0.9642 | 1.2104 | 0.9377 | 1.0035 | 1.0352 | 1.0092 |
| CMF | 0.8272 | 0.7977 | 0.8120 | 0.7945 | 0.7987 | 0.8088 |
| MCF | 0.8061 | 0.7914 | 0.7907 | 0.7761 | 0.7859 | 0.7913 |
| MCF-LF | **0.8017** | **0.7644** | **0.7806** | **0.7607** | **0.7504** | **0.7755** |

Table 2: Comparison of different methods on the Book-Crossing data. Each column records the RMSE scores on one domain and the last column records the RMSE score on the total testing data. Each row records the mean RMSE of the corresponding method over 10 trials. 1st domain: 'Mystery & Thrillers'; 2nd domain: 'Science Fiction & Fantasy'; 3rd domain: 'Science'; 4th domain: 'Business & Investing'; 5th domain: 'Religion & Spirituality'.

| Method | 1st domain | 2nd domain | 3rd domain | 4th domain | 5th domain | Total |
| --- | --- | --- | --- | --- | --- | --- |
| PMF | 0.9180 | 0.9795 | 0.8308 | 0.8699 | 0.8812 | 0.9269 |
| CMF | 0.9620 | 1.0207 | 0.9777 | 0.8465 | 1.0449 | 0.9960 |
| MCF | 0.7023 | 0.7046 | 0.7585 | 0.7555 | 0.7371 | 0.7158 |
| MCF-LF | **0.5686** | **0.5791** | **0.6047** | **0.6001** | **0.5953** | **0.5811** |

Table 4: Mean of correlation matrix learned by MCF-LF on the Book-Crossing data in different domains. 1st domain: 'Mystery & Thrillers'; 2nd domain: 'Science Fiction & Fantasy'; 3rd domain: 'Science'; 4th domain: 'Business & Investing'; 5th domain: 'Religion & Spirituality'.

|  | 1st | 2nd | 3rd | 4th | 5th |
| --- | --- | --- | --- | --- | --- |
| 1st | 1.0000 | 0.6839 | 0.4973 | 0.0137 | 0.3887 |
| 2nd | 0.6839 | 1.0000 | 0.4636 | -0.0489 | 0.6034 |
| 3rd | 0.4973 | 0.4636 | 1.0000 | 0.7270 | 0.7525 |
| 4th | 0.0137 | -0.0489 | 0.7270 | 1.0000 | 0.6682 |
| 5th | 0.3887 | 0.6034 | 0.7525 | 0.6682 | 1.0000 |

Another way to alleviate the data sparsity problem in CF is to apply active learning [3, 12]. Unlike many conventional machine learning methods which wait passively for labeled data to be provided in order to start the learning process, active learning takes a more active approach by selecting unlabeled data points to query some oracle or domain expert to reduce the labeling cost. For our future work, we are interested in incorporating active learning into our probabilistic model to further boost the learning performance.

## Acknowledgements

This research has been supported by General Research Fund 622209 from the Research Grants Council of Hong Kong.

## References

[1] C. Basu, H. Hirsh, and W. W. Cohen. Recommendation as classification: Using social and content-based information in recommendation. In *Proceedings of the 15th National Conference on Artificial Intelligence*, pages 714–720, Madison,Wis, USA, 1998.

[2] D. Billsus and M. J. Pazzani. Learning collaborative information filters. In *Proceedings of the 15th International Conference on Machine Learning*, pages 46–54, Madison, Wisconsin, USA, 1998.

[3] C. Boutilier, R. Zemel, and B. Marlin. Active collaborative filtering. In *Proceedings of the 19th Conference on Uncertainty in Artificial Intelligence*, pages 98–106, San Francisco, CA, 2003.

[4] J. Breese, D. Heckerman, and C. Kadie. Empirical analysis of predictive algorithms for collaborative filtering. In *Proceedings of the 14th Conference on Uncertainty in Artificial Intelligence*, pages 43–52, Madison, WI, 1998.

[5] D. DeCoste. Collaborative prediction using ensembles of maximum margin matrix factorizations. In *Proceedings of the 23rd International Conference on Machine Learning*, pages 249–256, Pittsburgh, Pennsylvania, USA, 2006.

[6] M. Fazel, H. Hindi, and S. Boyd. A rank minimization heuristic with application to minimum order system approximation. In *Proceedings American Control Conference*, pages 4734–4739, 2001.

[7] M. Fazel, H. Hindi, and S. Boyd. Log-det heuristic for matrix rank minimization with applications to hankel and euclidean distance matrices. In *Proceedings American Control Conference*, pages 2156–2162, 2003.

[8] A. K. Gupta and D. K. Nagar. *Matrix variate distributions*. Chapman & Hall, 2000.

[9] D. Heckerman, D. Chickering, C. Meek, R. Rounthwaite, and C. Kadie. Dependency networks for collaborative filtering and data visualization. In *Proceedings of the 16th Conference on Uncertainty in Artificial Intelligence*, pages 264–273, Stanford University, Stanford, California, USA, 2000.

[10] J. L. Herlocker, J. A. Konstan, A. Borchers, and J. Riedl. An algorithmic framework for performing collaborative filtering. In *Proceedings of the 22nd Annual International ACM SIGIR Conference on Research and Development in Information Retrieval*, pages 230–237, Berkeley, CA, USA, 1999.

[11] R. Jin, L. Si, and C. Zhai. Preference-based graphic models for collaborative filtering. In *Proceedings of the 19th Conference in Uncertainty in Artificial Intelligence*, pages 329–336, Acapulco, Mexico, 2003.

[12] Rong Jin and Luo Si. A bayesian approach toward active learning for collaborative filtering. In *Proceedings of the 20th Conference in Uncertainty in Artificial Intelligence*, pages 278–285, Banff, Canada, 2004.

[13] Y. Koren. Factorization meets the neighborhood: a multifaceted collaborative filtering model. In *Proceedings of the 14th ACM SIGKDD International Conference on Knowledge Discovery and Data Mining*, Las Vegas, Nevada, USA, 2008.

[14] N. D. Lawrence and R. Urtasun. Non-linear matrix factorization with gaussian processes. In *Proceedings of the 26th International Conference on Machine Learning*, pages 601–608, Montreal, Quebec, Canada, 2009.

[15] B. Li, Q. Yang, and X. Xue. Transfer learning for collaborative filtering via a rating-matrix generative model. In *Proceedings of the 26th International Conference on Machine Learning*, pages 617–624, Montreal, Quebec, Canada, 2009.

[16] M. R. McLaughlin and J. L. Herlocker. A collaborative filtering algorithm and evaluation metric that accurately model the user experience. In *Proceedings of the 27th Annual International ACM SIGIR Conference on Research and Development in Information Retrieval*, pages 329–336, Sheffield, UK, 2004.

[17] P. Melville, R. J. Mooney, and R. Nagarajan. Content-boosted collaborative filtering for improved recommendations. In *Proceedings of the 8th National Conference on Artificial intelligence*, pages 187–192, Edmonton, Alberta, Canada, 2002.

[18] D. Pennock, E. Horvitz, S. Lawrence, and C. Giles. Collaborative filtering by personality diagnosis: A hybrid memory- and model-based approach. In *Proceedings of the 16th Conference on Uncertainty in Artificial Intelligence*, pages 473–480, Stanford University, Stanford, California, USA, 2000.

[19] A. Popescul, L. Ungar, D. Pennock, and S. Lawrence. Probabilistic models for unified collaborative and content-based recommendation in sparse-data environments. In *Proceedings of the 17th Conference on Uncertainty in Artificial Intelligence*, pages 437–444, University of Washington, Seattle, Washington, USA, 2001.

[20] J. D. M. Rennie and N. Srebro. Fast maximum margin matrix factorization for collaborative prediction. In *Proceedings of the 22nd International Conference on Machine Learning*, pages 713–719, Bonn, Germany, 2005.

[21] R. Salakhutdinov and A. Mnih. Probabilistic matrix factorization. In J.C. Platt, D. Koller, Y. Singer, and S. Roweis, editors, *Advances in Neural Information Processing Systems 20*, pages 1257–1264, Vancouver, British Columbia, Canada, 2007.

[22] R. Salakhutdinov and A. Mnih. Bayesian probabilistic matrix factorization using Markov chain Monte Carlo. In *Proceedings of the 25th International Conference on Machine Learning*, pages 880–887, Helsinki, Finland, 2008.

[23] B. M. Sarwar, G. Karypis, J. A. Konstan, and J. Riedl. Item-based collaborative filtering recommendation algorithms. In *Proceedings of the 10th International World Wide Web Conference*, pages 285–295, Hong Kong, 2001.

[24] A. P. Singh and G. J. Gordon. Relational learning via collective matrix factorization. In *Proceedings of the 14th ACM SIGKDD International Conference on Knowledge Discovery and Data Mining*, pages 650–658, Las Vegas, Nevada, USA, 2008.

[25] N. Srebro, J. D. M. Rennie, and T. S. Jaakkola. Maximum-margin matrix factorization. In L. K. Saul, Y. Weiss, and L. Bottou, editors, *Advances in Neural Information Processing Systems 17*, pages 1329–1336, Vancouver, British Columbia, Canada, 2005.

[26] X. Su and T. M. Khoshgoftaar. A survey of collaborative filtering techniques. *Advances in Artificial Intelligence*, 2009.

[27] M. E. Tipping and C. M. Bishop. Probabilistic principal component analysis. *Journal of the Royal Statistic Society, B*, 61(3):611–622, 1999.

[28] T. Truyen, D. Phung, and S. Venkatesh. Ordinal boltzmann machines for collaborative filtering. In *Proceedings of the 25th Conference on Uncertainty in Artificial Intelligence*, pages 548–556, Corvallis, Oregon, 2009.

[29] K. Yu, J. D. Lafferty, S. Zhu, and Y. Gong. Large-scale collaborative prediction using a nonparametric random effects model. In *Proceedings of the 26th International Conference on Machine Learning*, pages 1185–1192, Montreal, Quebec, Canada, 2009.

[30] K. Yu, A. Schwaighofer, V. Tresp, W.-Y. Ma, and H. Zhang. Collaborative ensemble learning: Combining collaborative and content-based information filtering via hierarchical bayes. In *Proceedings of the 19th Conference on Uncertainty in Artificial Intelligence*, pages 616–623, Acapulco, Mexico, 2003.

[31] K. Yu, A. Schwaighofer, V. Tresp, X. Xu, and H.-P. Kriegel. Probabilistic memory-based collaborative filtering. *IEEE Transactions on Knowledge and Data Engeering*, 16(1):56–69, 2004.

[32] K. Yu, S. Zhu, J. D. Lafferty, and Y. Gong. Fast nonparametric matrix factorization for large-scale collaborative filtering. In *Proceedings of the 32nd Annual International ACM SIGIR Conference on Research and Development in Information Retrieval*, pages 211–218, Boston, MA, USA, 2009.